\documentclass[epj,final]{svjour}

\begin{document}

\authorrunning{T. Schneider, J. M. Singer}

\titlerunning{Fundamental constraints for the mechanism of superconductivity}

\title{Fundamental constraints for the mechanism of superconductivity in cuprates}

\author{T. Schneider and J. M. Singer}

\institute{Physikinstitut, Universit\"at Z\"urich, CH-8057 Z\"urich, Switzerland}

\mail{J. M. Singer, Physikinstitut, Universit\"at Z\"urich, Winterthurerstr. 190, CH-8057 Z\"urich, Switzerland,
\email{jms@physik.unizh.ch}}

\date{Received: \today / Revised version: date}

\abstract{
Considerable progress has been made over the last decade in understanding
the phenomenological properties of the cuprate high-T$_{c}$ superconductors
and in producing well characterized high quality materials. Nevertheless,
the pairing mechanism itself remains controversial.
We establish a criterion to test theories for layered superconductors
relying on a substantial interlayer contribution. The criterion is based on
the ratio of the interlayer contribution to the total superfluid density,
which is traced back to the inverse squared effective mass anisotropy, 
${1/(1+2\gamma^{2})}$. 
$\gamma$ can be measured rather accurately by various
experimental techniques. It turns out that models relying on interlayer
pairing cannot be considered as serious candidates for the mechanism of
superconductivity in cuprate superconductors.}


\maketitle

One candidate mechanism to explain superconductivity in the cuprates 
is the interlayer tunneling (ILT) model proposed by
P. W. Anderson and coworkers 
\cite{Charkravarty}-\cite{Anderson2}.
There, superconductivity is supposed to result primarily from an interlayer
coupling mechanism. It has been argued \cite{Anderson2,Leggett}, that the
comparison of the measured interlayer magnetic 
penetration depth $\lambda _{c}$ with the
value determined from the ILT-model condensation energy, $\lambda_{c}^{{\rm %
ILT}}$, is a crucial test (c denotes the c-axis of the unit cell). 
Recent direct measurements of $\lambda_{c}$ in $%
{\rm Tl_{2}Ba_{2}CuO_{6+\delta }}$ \cite{Moler,Marel} and ${\rm %
HgBa_{2}CuO_{4+\delta }}$ \cite{Kirtley} make it unlikely that the present
version of the ILT model is a serious candidate for the mechanism of
superconductivity in cuprate superconductors. Indeed, $\lambda _{c}$ turns
out to be much larger than $\lambda _{c}^{{\rm ILT}}$ 
\cite{Anderson2}-\cite{Kirtley}.
However, it is important to
recognize, that this approach is only applicable to theories, which can
provide an estimate for $\lambda _{c}$.

Here we introduce a more general measure for the ratio between the
interlayer - and total pairing interaction in the superconducting state. For
this purpose the system is subjected to phase twists $k_{i}$ along three
respective crystallographic axes $i=a,b,c$. In the presence of such a phase
twist and in the limit $k_{i}\rightarrow 0$, the free energy density then
reads as (see e.g. \cite{Fisher,Kim}) 
\begin{equation}
f_{i}=\frac{k_{B}T}{2}\Upsilon _{i}k_{i}^{2}\quad .
\end{equation}
The helicity modulus $\Upsilon _{i}$ is given as 
\begin{equation}
\Upsilon _{i}=\frac{\Phi _{0}^{2}}{16\pi ^{3}\lambda _{i}^{2}}\quad ,
\end{equation}
\begin{equation}
\frac{1}{\lambda _{i}^{2}}=\frac{16\pi ^{3}\hbar ^{2}n_{s}}{M_{i}\Phi
_{o}^{2}}\quad .
\end{equation}
$\Phi _{0}$ is the flux quantum, $\lambda _{i}$ the magnetic penetration
depth, $M_{i}$ denotes the effective pair mass appearing in the gradient term
of an anisotropic Ginzburg-Landau action and $n_{s}$ is the superfluid
number density. Imposing such twists of magnitude $|k_{i}|$ along the
directions $i=a,b,c$, respectively, the ratio 
\begin{eqnarray}
\eta &=& \frac{f_{c}}{f_{a}+f_{b}+f_{c}}=\frac{\Upsilon _{c}}{\Upsilon
_{a}+\Upsilon _{b}+\Upsilon _{c}}\cr
\cr
&=& 1/\left( ({\lambda _{c}^{2}\left(1/ \lambda
_{a}^{2}+1/\lambda _{b}^{2}+1/\lambda _{c}^{2}\right) }\right)  
\end{eqnarray}
measures the fraction which the interlayer coupling contributes to the total
free energy density of the superfluid. For tetragonal systems, where 
$\lambda _{a}=\lambda _{b}\equiv \lambda _{\parallel }$, 
$\lambda _{c}\equiv \lambda _{\perp }$, it reduces
to the simple form 
\begin{equation}
\eta =\frac{1}{1+2\gamma ^{2}}\quad ,\quad \gamma =\sqrt{\frac{M_{\perp }}{%
M_{\parallel }}}\quad .
\end{equation}
$\gamma $ is the anisotropy parameter, which can be measured by various
experimental techniques. $\eta =1/3$ corresponds to the case where the
pairing interaction supplies the same fraction
to the superfluid in the a-, b- and c-direction,
while in a two-dimensional superconductor, $\eta =0.$ In Table \ref{table1}
we list some experimental estimates for cuprate superconductors close to
optimum doping \ and -- for comparison -- we included the conventional
layered superconductor $\mathrm{NbSe_{2}}$. 

{\begin{table}
{\begin{center}
\begin{tabular}{|l|c|c|c|c|}
\hline
& $T_c\ [K]$ & $\gamma$ & $\eta$ & Source \\ \hline
${\rm YBa_2Cu_3O_{7-\delta}}$ & 91.7 & 8.95 & 0.006 & \cite{Chien,Schneider} \\
${\rm La_{2-x}Sr_{x}CuO_4}$ & 35 & 14.0 & 0.0025 & \cite{Maeda} \\
${\rm HgBa_2CuO_{4.1}}$ & 94.1 & 26.7 & 0.0007 & \cite{Hofer} \\
${\rm HgBa_2Ca_2Cu_3O_{8+\delta}}$ & 133 & 52 & 0.0002 & \cite{Vulcanescu} \\
${\rm Tl_2Ba_2CuO_{6+\delta}}$ & 87.6 & 117 & 0.00004 & \cite{Zuo}
\\ \hline
${\rm NbSe_2}$ & 7 & 3 & 0.053 & \cite{Hussey} \\ \hline
\end{tabular}
\end{center}}
\caption{Various experimental estimates for $\protect\gamma $ and $\protect%
\eta $.}
\label{table1}
\end{table}
}

Noting that optimally doped $\mathrm{YBa_{2}Cu_{3}O_{7-\delta }}$ is the most
isotropic cuprate and  $\gamma $ is known to increase by approaching the
underdoped limit \cite{Chien,Hofer,Hubbard}, the listed  $\eta $ values
clearly reveal  that superfluidity in cuprate superconductors is nearly two
- dimensional. Consequently, the small interlayer pairing contribution to
the superfluid makes it unlikely that theoretical models relying on a
significant interlayer pairing contribution, such as the ILT model, are
serious candidates for the mechanism of superconductivity in the cuprates. \
As $\mathrm{HgBa_{2}CuO_{4+\delta }}$ and $\mathrm{Tl_{2}Ba_{2}CuO_{6+\delta }}$
are concerned, our results are consistent with previous estimates 
\cite{Moler}-\cite{Kirtley}, but our approach does not rely on a particular type of
$\lambda _{c}$ measurement and a model dependent estimate of this quantity, but more
generally on determination of the anisotropy $\gamma $. 
This quantity can be deduced with rather high precision from magnetization \cite{Hubbard},
specific heat \cite{Roulin}, magnetic torque \cite{Schneider}, etc.
measurements on bulk samples. To summarize, we have shown that models
relying on interlayer coupling cannot be considered as candidates for the
mechanism of superconductivity in cuprate superconductors. Indeed the
interlayer contribution to the superfluid is
very small. For this reason the materials can be viewed as a stack of weakly
coupled superconducting slabs of finite thickness.

\end{document}